\documentclass[aps,pra,twocolumn,showpacs,preprintnumbers]{revtex4-1}

\usepackage{psfrag,graphicx}
\usepackage{dcolumn}
\usepackage{amsmath,amssymb}
\usepackage{bm}
\usepackage{amsfonts,amssymb,amsmath}        
\usepackage{epstopdf}

\newcommand{\be}{\begin{equation}}
\newcommand{\ee}{\end{equation}}
\newcommand{\bq}{\begin{eqnarray}}
\newcommand{\eq}{\end{eqnarray}}

\newcommand{\ket}[1]{\left | \, #1 \right\rangle}

\begin{document}

\title{A witness for topological order and stable quantum memories in abelian anyonic systems}
\author{James R. Wootton}
\affiliation{Department of Physics, University of Basel, Klingelbergstrasse 82, CH-4056 Basel, Switzerland\\ School of Physics and Astronomy, University of Leeds, Leeds LS2 9JT,U.K.}

\begin{abstract}

We propose a novel parameter, the anyonic topological entropy, designed to detect the error correcting phase of a topological memory. Unlike similar quantities such as the topological entropy, the anyonic topological entropy is defined using the states of the anyon occupations. As such, though the parameter deals with phases and phase transitions that are quantum in nature, it can be calculated solely from classical probability distributions. In many cases, these calculations will be tractable using efficient classical algorithms. The parameter therefore provides a new avenue for efficient studies of anyonic systems.

\end{abstract}


\maketitle

\section{Introduction}

Topological models that support anyonic quasiparticles have generated a great deal of interest, both in condensed matter physics \cite{wen,honeycomb,stringnet,hamma} and quantum information theory. This is due in the most part to proposals for fault-tolerant quantum computation \cite{preskill,double,dennis,freedman,raussendorf,pachos,wootton}. The topologically ordered states of these models cannot, in general, be detected by any local order parameter. Instead, non-local order parameters can be used, such as the topological entanglement entropy \cite{kitpres,levwen,woottonent,hastings}. Though these are well suited to many problems, their calculation can be difficult in some cases. Also, some are defined specifically for pure states, and can run into problems when mixed states are considered \cite{woottonent}. As such, a wide toolkit of parameters are desirable, so that a wide variety of problems can be tackled.

Here we propose a novel parameter that is explicitly defined to provide a witness for both error correctability and the presence of topological order in models with anyonic quasiparticles. As such, we precede the definition of the parameter with a discussion of these issues and the conditions they place upon a system.

\section{Topological quantum memories and error correction}

\begin{figure}[t]
\begin{center}
{\includegraphics[width=8.5cm]{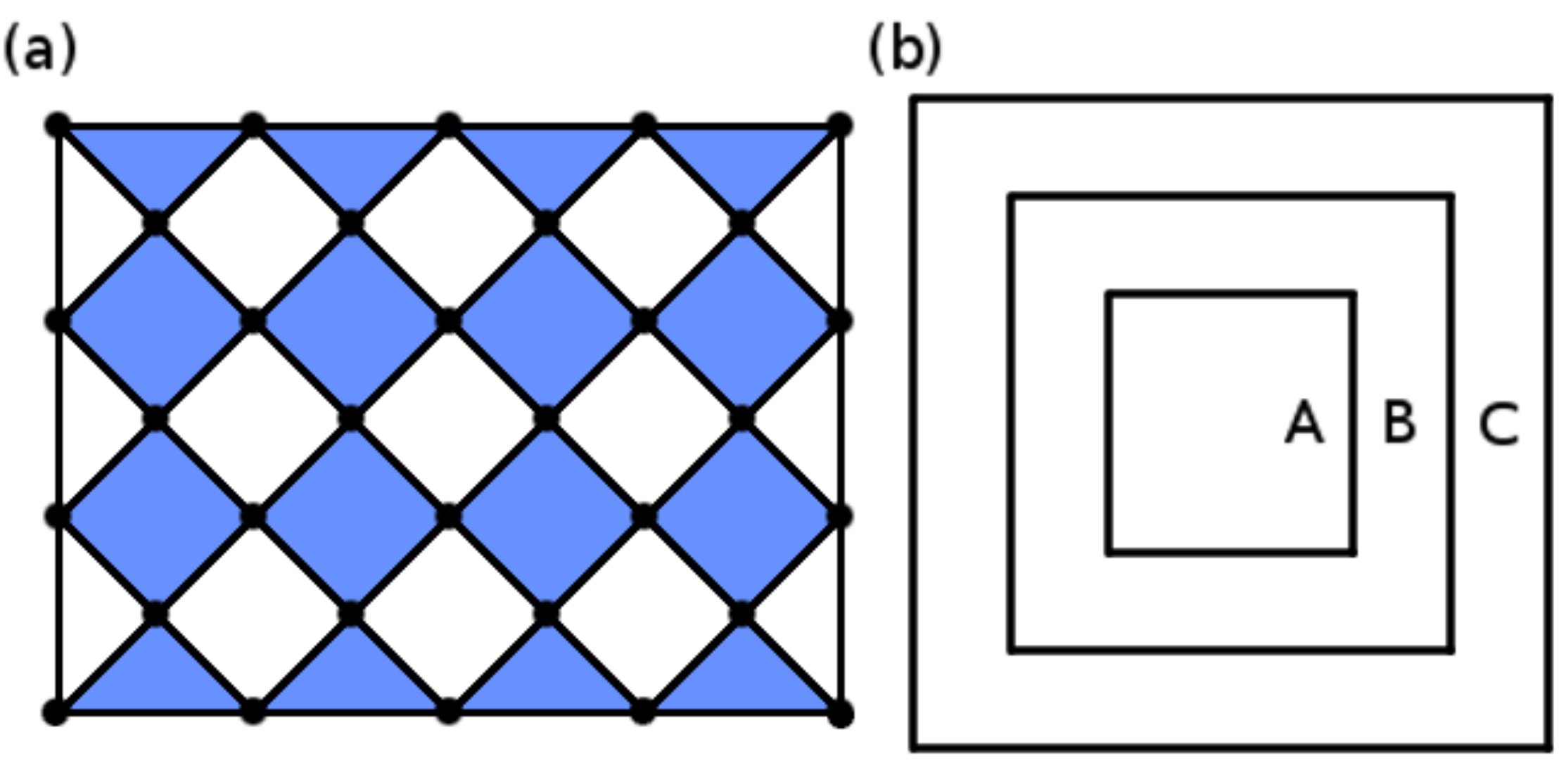}}
\caption{\label{fig1} (a) The spin lattice on which an $L \times L$ planar code is defined, with $s$-plaquettes shown in blue and $p$-plaquettes in white. A spin-$1/2$ particle resides on each vertex. The linear size $L$ is characterized the number of $s$-plaquettes along each side, with $L=4$ in this case.  (b) The regions used for the calculation of the anyonic topological entropy. Note that these partitions are of anyonic pseudospins, not of physical spins. The thickness of the region $B$ must be on the order of its height, and all dimensions should be on the order of the system size.}
\end{center}
\end{figure}

For any topological model that can be used as a quantum memory, it is important to know where the boundaries of the error correcting phase lie. To see how this may be done, let us consider the planar code \cite{double,dennis} as a concrete example. This is defined on the spin lattice of Fig. \ref{fig1}(a), where a spin-$1/2$ particle is placed on each vertex and the plaquettes are bicoloured as $p$- and $s$-plaquettes. The following Hermitian operators are then defined around each of these,
\be
A_s = \prod_{i \in s} \sigma^x_i, \,\,\, B_p = \prod_{i \in p} \sigma^z_i.
\ee
These operators determine the anyonic occupation of their corresponding plaquettes. Any state within the $-1$ eigenspace of an operator is said to have an anyon residing on the corresponding plaquette. These are so-called magnetic anyons $m$ on the $p$-plaquettes and electric anyons $e$ on the $s$-plaquettes. A pair of $e$ anyons is created on neighbouring plaquettes whenever a $\sigma^z$ operator is applied to their shared spin. Further applications of $\sigma^z$ operators on spins forming a chain can be used to move the anyons around the lattice. Similarly a pairs of $m$ anyons are created and moved by $\sigma^x$ operations.

Since all the $A_s$ and $B_p$ operators commute, they form the stabilizers of a stabilizer code. The anyonic vacuum, the subspace of states for which no anyons are present on any plaquette, is the corresponding stabilizer space. For the planar code, this space is two-dimensional, allowing a single qubit to be stored. These two states correspond to the anyonic occupations of the edges, which are not determined by the stabilizers. The left and right edges each hold either the vacuum or an $e$ anyon, and the top and bottom hold either vacuum or an $m$. The $X$ ($Z$) basis of the stored qubit may be chosen such that the $\ket{+}$ ($\ket{0}$) state corresponds to the vacuum on the top (left) edge and $\ket{-}$ ($\ket{1}$) corresponds to an $m$ ($e$) anyon.

The effect of errors, which occur due to perturbations, thermal effects or other unwanted influences, is to move anyons, or to create and annihilate them in pairs. If the anyons are moved off of the edges of the code, logical errors occur on the stored information. With the conventions chosen above, a logical $X$ is caused if an $m$ anyon is moved off of the top edge, and a logical $Z$ when an $e$ is moved off of the left edge. When the errors are sufficiently weak, it is possible to perform a correction procedure to undo the logical errors caused. Such a procedure involves measuring the anyonic occupation of each plaquette and, using that information, deducing the parity of the number of $e$ anyons that were moved off of the top edge. If the number is odd then the net effect is that a single $e$ has been moved off the edge, and so a logical $X$ error is caused. If the number is even then there is no net anyon, since an even number annihilate to the vacuum, and so no logical error. Similarly, the parity of the number of $m$'s moving off of the left edge can be determined to see whether a logical $Z$ error has occurred. Once it is known what logical errors have been caused, they can be undone, and the stored information retrieved reliably. However, when the errors are sufficiently strong, such a correction procedure is no longer possible. The distribution of anyons becomes such that there is no reliable means to determine the nature of net anyon that has moved across any edge, and so whether logical errors have occurred. 

Successful error correction therefore requires there to exist an algorithm that, given the anyonic occupancies on the code, can determine the net anyon that has been moved over each edge. A more general statement of this condition is required for more complex encoding schemes, such as that in \cite{raussendorf}. In general, successful error correction requires there to exist an algorithm that, given the anyonic occupancies of a region bounded by multiple independent edges, can determine the net anyon that has been moved over each edge.

\section{Topological entropy and loop correlations}\label{loop}

To determine whether the state of an anyonic system belongs to the topologically ordered phase, the topological entropy is usually used as an order parameter \cite{kitpres,levwen}. This entropy is defined such that it detects whether loop correlations that are present in the state of a system. These are non-local correlations along closed loops that do not exist along open strings. They are present in the topologically ordered states of anyonic systems, and absent in topologically trivial states.

Let us consider the topological entropy of Levin and Wen \cite{levwen,woottonent}, and review its definition and relation to loop correlations. For this quantity an annulus shaped region of the system is considered and split into three. The region $A$ is the bottom of the annulus, $B$ is the sides and $C$ is the top. The topological entropy is then,
\be
\gamma = I_{A,BC} - I_{A,B}.
\ee
Here $I_{A,B} = S_A + S_B - S_AB$ is the mutual information between the regions $A$ and $B$, and $S_A$ is the Von Neuman, entropy of the region $A$, etc. This entropy therefore quantifies the correlations that can be detected only when the entire annulus is considered, and is not detectable for any horseshoe shaped subregion. Put another way, it quantifies the correlations that exist only around closed loops, and not open strings.

It is intructive to consider the a stabilizer state of the planar code as a concrete example. Let us use $\pi$ to denote the set of all plaquettes with full or partial support on the spins enclosed by the annulus, and $\sigma$ to denote the corresponding set of vertices. The loop operators may then be defined,
\be
L_\sigma = \prod_{s \in \sigma} A_s, \,\,\, L_\pi = \prod_{p \in \pi} B_p.
\ee
The action of these operators on all spins enlosed by the annulus will cancel, leaving only non-trivial support on the annulus. We call these $s$- and $p$-type loop correlation functions, respectively. The $L_\sigma$ ($L_\pi$) operator will have eigenvalue $+1$ in any stabilizer state, and so an even number of spins around the loop will be in state $\ket{-}$ ($\ket{0}$). The parity of the numbers of these states in $A$ must then be equal to that in $BC$, but no such condition is present for $A$ and $B$ alone. These correlations therefore contribute $\ln 2$ for each loop in the first term of $\gamma$ and nothing to the second. All other correlations are local, and contribute equally to each. This leads to $\gamma = 2 \ln 2$ for this state.

Note that the $L_\sigma$ and $L_\pi$ operators, as products of individual $A_s$ and $B_p$, determine the net occupation of the $\sigma$ vertices and $\pi$ plaquettes. Net vacuum or $e$ ($m$) correponds to the $+1$ and $-1$ eigenspaces of $L_\sigma$ ($L_\pi$), respectively. Any state that has a definite net anyon occupation for the enclosed region (like the stabilizer space) is therefore an eigenstate of these operators. It will therefore have loop correlations that lead to a topological entropy of $\gamma = 2 \ln 2$, meaning such states are topologically ordered. 

The situation becomes more complex when the net occupancy of the enclosed region is not definite. For example, consider the application of single spin $\sigma^x$ and $\sigma^z$ to a state within the stabilizer space, with each error occurring with probability $P$. For $P>0$ this will yield a mixed state of multiple anyon configurations. If the loops are large, the total anyon occupancies of the region enclosed by the loop will be completely random. The probability of an odd or even number of $\ket{1}$'s around the $L_{\pi}$ loop will be equal, and so there will be no correlations in this between $A$ and $BC$. However, this does not mean that the loop correlations have disapeared, only that the means for their detection must be changed by modifying the loop correlation function. To see why and how, consider measurement of the anyon configuration on an annulus surrounding the loop. If $P$ is sufficiently small, and the annulus is sufficiently thick, an algorithm like that use for error correction can be used to determine the net anyon that has been moved over the inner edge of the annulus by the errors. Since this edge completely encloses this region, this therefore determines the net occupation of the enclosed region. Once its net occupation becomes definite, the resulting state is an eigenstate of the loop correlation functions and hence topologically ordered. If $P$ is too high, however, the net anyon occupancy of the enclosed region can no longer be determined from the anyons of the annulus, and hence remains random even after the measurement. This shows that it is no longer topologically ordered.

Rather than actually performing such a measurement, it may be incorporated into `thickened' loop correlation functions with support over the entire annulus \cite{levwen,hastings}. These are defined such that a state is an eigenstate if the total anyon occupation of the region enclosed by the loop, and hence by the annulus, can be determined by the anyon configuration on the annulus. The existence of such thickened loop operators means that loop correlations exist, and hence the topological entropy will take a non-zero value. The state is therefore topologically ordered. This is true not just for the planar code, but in general \cite{hastings}.

This possibility gives us a novel criterion for assessing whether loop correlations are present for states of models that support anyonic quasiparticles. If, by considering the anyon configuration on an annulus, we can determine the total anyon occupancy (both $e$ and $m$) enclosed the annulus, we know that both $p$- and $s$-type loop correlation functions can be defined for which the state is an eigenstate. It therefore has loop correlations, a non-zero topological entropy and will be topologically ordered.

If the state is an eigenstate of only one type of loop operator, and hence only the net $e$ or $m$ occupancy within the annulus can be determined but not both, then there is some ambiguity. It could be that the `classically topologically ordered' states of \cite{claudios} are present. However, it may also be that the state is not only an eigenstate of loop operators, but also of open string operators. It will therefore not have loop correlations or topological order. For example, we can consider the spin-polarized state with all spins in state $\ket{0}$. This is obviously an eigenstate of large $p$-type loop operators in the planar code, since it corresponds to a complete lack of $m$ anyons. However, it is similarly an eigenstate of open $\sigma^z$ strings. Clearly this state is topological trivial.

The reason why this ambiguity does not exist when the state is an eigenstate of both types of loop operators is because any incomplete part of a $p$-type loop, which will form an open string, will anticommute with $s$-type loops, and vice-versa. As such, states that are an eigenstate of open $p$-type strings will not be an eigenstate of $s$-type loops, and vice-versa. A state that is an eigenstate of both $p$- and $s$-type loops therefore is not an eigenstate of either type of open string operators, and hence must truly have loop correlations and topological order. The same argument holds for all other Abelian quantum double models \cite{double}, and can be expected to hold for Abelian anyon models in general.

\section{Defining the anyonic topological entropy}

Note that the criterion for loop correlations parallels the condition for successful error correction in topological memories as discussed above. The existence of the algorithm required for error correction directly implies that thickened loop operators can indeed be defined, and hence that the state is an eigenstate of loop correlation functions. To see this, consider the annulus formed by the region $B$, depicted in Fig. \ref{fig1}(b). This has two edges, one outer and one inner. The inner edge completely encompasses the region $A$. As such, if there exists an algorithm that can determine the net anyon moved over each independent edge of a bounded region, it may be applied to the annulus to determine the net anyonic occupation of the region $A$ using the anyonic occupancies of $B$. A thickened loop operator can then be defined on the annulus for which the state is an eigenstate.

These properties of error correction and loop correlations allow a quantity to be defined, whose value signals whether the former is possible and the latter is present. To define this quantity, first note that any state of the underlying spin lattice can be completely described by specifying the anyonic occupancies of all plaquettes, as well as the internal state of the quantum memory \cite{iblisdir}. Rather than considering the physical spins of the system we may therefore map the model to one of anyonic pseudospins. One of these is assigned to each plaquette, with the states $\ket{0}$ and $\ket{1}$ used to denote that the corresponding plaquette is occupied by the vacuum or an anyon, respectively. A final pseudospin, representing the state of the quantum memory, can be defined in terms of the occupation of the left or top edges. However this will not enter into the arguments below.

Consider a partition of the anyonic pseudospins according to the annulus of Fig. \ref{fig1} (b). Using this, $\rho_B$ can be used to denote the reduced density matrix of all anyonic pseudospins within the region $B$, obtained from the state of the entire system by tracing out all other pseudospins. Similarly, $\rho_{\tilde A B}$ can be used to denote the joint density matrix for the anyonic pseudospins of region $B$ and the net anyon occupation of region $A$. This is obtained from the state of the entire system by tracing out all pseudospins not in $A$ or $B$, and then also tracing out all degrees of freedom within $A$ except that of the net occupation. The entropies corresponding to these density matrices, $S_B = -\rho_B \log \rho_B$ and $S_{\tilde A B} = -\rho_{\tilde A B} \log \rho_{\tilde A B}$, can then be used to define the following quantity,
\be
\Gamma' = S_{\tilde{A}B}-S_B.
\ee
This quantifies the information concerning the net occupation of $A$ that cannot be deduced from the anyonic occupation of $B$. In the error correcting phase and as $L \rightarrow \infty$, knowledge of $B$ allows the net occupancy of $A$ to be deduced completely. As such, $\Gamma' = 0$. Since this value signifies the ability to completely deduce the anyon configuration within the annulus it is a witness, and a definite smoking gun, for the presence of topological order.

The uncorrectable phase is that for which logical $X$ and $Z$ errors both occur completely randomly and there is no way to use the anyon configuration of the bulk to deduce whether they have occurred or not. The former implies that the net anyon moved over any edge is completely random. The net anyon occupation of the region $A$ is therefore a random mixture of the four possibilities (vacuum, $e$ only, $m$ only and both $e$ and $m$), and so has an entropy of $S_{\tilde{A}B} = 2 \log 2$. The latter implies that the net occupation cannot be deduced from the anyon configuration of $B$, and so $S_{\tilde{A}B} = S_{\tilde{A}} + S_{B}$. As such $\Gamma' = S_{\tilde{A}} = 2 \log 2$ for this case. 

When $\Gamma'$ takes an intermediate value between these two extremes, we find the ambiguous case mentioned in Section \ref{loop}. In the thermodynamic limit, and assuming local interactions, the only possible intermediate value will be $\Gamma' = \log 2$, corresponding to unambiguous determination of the net occupation of only one anyon type. This may or may not signal that one type of loop correlation is present, depending on whether the underlying spin state is an eigenstate of open string operators as well as closed loops. In order to determine which is the case, additional insight into the model considered must be used. Otherwise, this intermediate value is neither a witness that topological order is present, or proof that it is not.

The quantity $\Gamma'$ is atypical in that it assigns the value of zero to states that are ordered and a value of $2 \log 2$ to those that are not. As such let us define and use the following modified version,
\be
\Gamma = 2 \log 2 - \Gamma' = 2 \log 2  + S_B - S_{\tilde{A}B}.
\ee
For this, it is the value $\Gamma = 2 \log 2$ that is the witness of states in the topologically ordered and error correcting regime (corresponding to $\Gamma'=0$), and $\Gamma = 0$ for those that are not (corresponding to $\Gamma'=2 \log 2$). It is this quantity that we call the anyonic topological entropy.

It is important to note that, though $\Gamma = 0$ implies error correction is not possible, this only holds for the particular definitions of the anyons used in the calculations. There may still be a definition of anyons for which $\Gamma = 2 \log 2$, and so error correction can be performed. For example, a state stabilizer by all the $A_s$ and $B_p$ operators of the planar code is clearly topologically ordered. When the anyonic occupations are defined by the eigenspaces of these operators, this state corresponds to the anyonic vacuum and gives $\Gamma = 2 \log 2$. However, if the occupations were instead defined by rotated operators $A'_s$ and $B'_p$, for which each $\sigma^x$ is replaced with a $\sigma^z$ and vice-versa, the distribution of anyons would appear completely random and uncorrectable, giving $\Gamma = 0$.

Note that these arguments hold when the distribution of anyons is classical, but could run into complications for a quantum distribution (since, for example, it is possible for $S_B$ to be greater than $S_{\tilde{A}B} $). However, since error correction depends only on correlations in the anyonic basis, calculations can be made for quantum states by removing the off-diagonal elements of the density matrices in this basis. The quantum states thus become classical, and the above may be applied normally.

Since the definition of the anyonic topological entropy is based on classical distributions of anyon occupancies, ignoring the coherence of the underlying spins, it gives a significant practical advantage over other methods in some cases. It allows the calculations of $\Gamma$, or al least bounds on the value, to be computed using efficient classical algorithms, such as those already used for error correction \cite{blossom,beat,montreal,fowler,woottoncorrect}.

\subsection{Relation to error correction}

The value of $\Gamma$ is discontinuous at phase transitions between topologically ordered and non-topologically ordered states. This discontinuity can be seen in the existing results of error correction algorithms \cite{beat,montreal,fowler,woottoncorrect}. Though these do not consider an annulus, they equivalently must determine the net anyon moved off independent edges of a bounded region. In these the probability of successful correction is found to be a step function in the thermodynamic limit, shifting from a value of $1$ below a threshold to $1/2$ above. This then becomes a step function between $2 \log 2$ and $0$ for $\Gamma$. This is considered in more detail in the Section \ref{corr}.

Since the definition of the anyonic topological entropy is based on error correction in topological memories, one might conclude that it is equivalent to a direct analysis of the fidelity of a quantum memory (achieved by performing error correction and determining with what rate this is successful in correcting errors). However, though they are indeed equivalent in all cases to which both may be applied, they have a very distinct difference which allows anyonic topological entropy to be applied to a wider variety of problems. The fidelity of a memory is a process based quantity: it only makes sense if the memory is prepared in an initial state to which an error model is then applied, such that the final state after error correction can be compared to the initial state. The anyonic topological entropy, however, is purely state based in its definition. It can be applied to any state, whether it is the result of a dynamic process (such as that considered by the fidelity), or whether it represents an equilibrium for which there is no concept of an error model or initial and final states. The anyonic topological entropy can then, for example, be applied to thermal states to determine whether a memory will be stable against finite temperature. An example of this is considered in section \ref{temp}.

\subsection{Relation to topological entropy}

The anyonic topological entropy, like the topological entropy, is a witness to the non-local loop correlations of topologically ordered states. In this sense they are equivalent. As described in Sec. \ref{loop}, if the net $e$ ($m$) anyon occupation of $A$ can be determined by the anyon configuration of $B$, then a modified $p$-type ($s$-type) loop operator can be defined for which the state is an eigenstate. This shows that the loop correlations exist, with each giving a contribution of $\log 2$ to the topological entropy $\gamma$ (here we use the definition of \cite{levwen}). Though this implies that $\Gamma = \gamma$ (at least when the correct definition for anyons is used for the former), this relation must be used with care. The ambiguities described above mean that the only definite case is for $\Gamma = 2 \log 2$. Here we know that both types of loop correlations are present are definitely present, and so $\gamma = 2 \log 2$. Otherwise $\Gamma$ and $\gamma$ only take the same value when ambiguities are not present.

\subsection{Generalization to other Abelian models}

The generalization of $\Gamma$ to other Abelian quantum double models is straightforward. One simply replaces the two level pseudo-spins used in the case of the planar code with ones of sufficient dimensionality to record the occupancies for the quasiparticles of the model. For example, for a quantum double model $D(G)$ based on an Abelian group $G$, there are $|G|$ different anyon types (including the vacuum) that can reside on each plaquette \cite{double}. Pseudo-spins of dimension $|G|$ are therefore required to record this information. The partitions and mutual informations are then applied exactly as above. In general for Abelian models, the total number of anyon types is given by $D^2$, where $D$ is the so-called quantum dimension of the model. The definition of $\Gamma$ then becomes,
\be
\Gamma = 2 \log D + S_B - S_{\tilde{A}B}.
\ee
The value of $\Gamma = 2 \log D$ is again the witness of topological order, since only topologically ordered states can realize the full confinement of anyons required for error correction. This value implies that the topological entropy will also be $\gamma = 2 \log D$. $\Gamma = 0$ again signifies that topological order is not present, and intermediate values are ambiguous. Generalization of the quantity to models with non-Abelian anyons or non-anyonic topological defects is not straightforward, but can be expected to follow from the same basic principles.


\section{Calculating $\Gamma$ using error correction} \label{corr}

As discussed above, given knowledge of the error model applied to an anyonic vacuum state, and given the anyon configuration within the bulk, an error correction algorithm can determine the most likely net anyon moved off each edge. If applied instead on the annulus of Fig. \ref{fig1}(b), this means that the algorithm can determine the most likely net occupation of the region $A$ when given the anyon configuration of $B$.

Using such an error correction algorithm, we can use numerical simulations to calculate $\Gamma$. Here we demonstrate how this may be done for the planar code subject independent bit and phase errors, where the $e$ and $m$ anyons can be considered separately. Corresponding calculations for other error models and topological codes can be done in a corresponding manner.

Let us use $P_s$ to denote the probability that the algorithm correctly guesses the net $e$ occupation of $A$, and $\pi_s$ to denote the probability that the net $e$ occupation of $A$ is the vacuum. Also we use $P_s' = 1-P_s$ and $\pi_s' = 1-\pi_s$. These values can be obtained numerically by running the error correction procedure for a large number of samples \cite{beat,montreal,fowler,woottoncorrect}. In terms of these quantities, the mutual information between the net $e$ occupation of $A$ and the guess of $B$ can be expressed,
\be \label{bound1}
I^s_{\tilde{A},B} = S(P_s \pi_s + P_s' \pi_s') - S(P_s).
\ee
Here $S(x) = -x \log x - (1-x) \log(1-x)$ denotes a Shannon entropy. Note that this value for $I^s_{\tilde{A},B}$ actually yields a lower bound, because error correction algorithms are not perfect. They cannot, in general, deduce as much information about the net occupation of $A$ as would be possible via brute force methods. A similar lower bound $I^p_{\tilde{A},B}$ can be calculated for the mutual information between the net $m$ occupation of $A$ and the guess of $B$. The lower bound for the total mutual information $I_{\tilde{A},B}$ is the sum of these two contributions.

Since $\Gamma'$ may be expressed $\Gamma' = S_{\tilde{A}} - I_{\tilde{A},B}$, the above quantity may then be used to lower bound $\Gamma$,
\be \label{bound2}
\Gamma \geq 2 \log 2 + I^s_{\tilde{A},B} + I^p_{\tilde{A},B} - S(\pi_s) - S(\pi_p).
\ee
This allows studies to be made of how $\Gamma$ changes with error rate, or with the time spent coupled to a thermal bath, that correspond to the results already obtained for error correction \cite{beat,montreal,fowler,woottoncorrect}. In these works, $P_s$ and $P_p$ approach a step function as $L \rightarrow \infty$, with $P_{s,p} \rightarrow 1$ in the error correcting regime and $P_{s,p} \rightarrow 1/2$ in the uncorrectable regime. The same would be true for brute force methods, with the only difference being that the imperfect algorithms have lower thresholds and take longer to reach their asymptotic values \cite{dennis}. From Eq.'s (\ref{bound1}) and (\ref{bound2}) we see that this will lead also to a step function in $\Gamma$, with $\Gamma = 2 \ln 2$ in the error correcting regime and $\Gamma = 0$ in the uncorrectable regime.

\section{Application to thermal states} \label{temp}

The topological entanglement entropy of thermal states of the toric code have been studied previously, though not without somewhat complex calculations \cite{claudios,iblisdir}. Here we demonstrate that corresponding studies of the planar code using the anyonic topological entropy are much simpler.

A study of thermal states first requires a Hamiltonian. The Hamiltonian $H$ of the planar code, and its corresponding form for the anyonic pseudo-spins, is,
\bq
H = - J_S \sum_s A_s - J_P \sum_p B_p \\
H_P = - J_S \sum_s \sigma^z_s - J_P \sum_p \sigma^z_p.
\eq
Given this Hamiltonian, the thermal state is the tensor product of thermal states for each individual pseudospin,
\be
\rho_s = \frac{e^{-J_S \beta \sigma^z_s}}{{\rm tr}(e^{-J_S \beta \sigma^z_s})}, \,\,\,\,\, \rho_p = \frac{e^{-J_P \beta \sigma^z_p}}{{\rm tr}(e^{-J_P \beta \sigma^z_p})},
\ee
where $\beta$ is the inverse temperature. The probability that an anyon exists on any plaquette can then be determined from these distributions to be $p_s = (1+\exp[2 J_S \beta])^{-1}$ and $p_p = (1+\exp[2 J_P \beta])^{-1}$.

Due to the fact that each plaquette's anyonic occupation is uncorrelated to any other, it is easy to see that $I_{\tilde{A},B}=0$. As such $\Gamma = 2 \log 2 - S_{\tilde{A}} = 2 \log 2 - S(\pi_S) - S(\pi_P)$, where $\pi_S$ ($\pi_P$) is the probability of an even parity of anyons, and hence a net occupation of the vacuum, in the $s$-plaquettes ($p$-plaquettes) of region $A$. These can be easily found from the probabilities $p_s$ and $p_p$ to be $\pi_S = (1+[1-2p_s]^{n_s})/2$ and $\pi_P = (1+[1-2p_p]^{n_p})/2$, where $n_s$ and $n_p$ are the number of $s$ and $p$ plaquettes in the region $A$, respectively.

\begin{figure}[t]
\begin{center}
{\includegraphics[width=8.5cm]{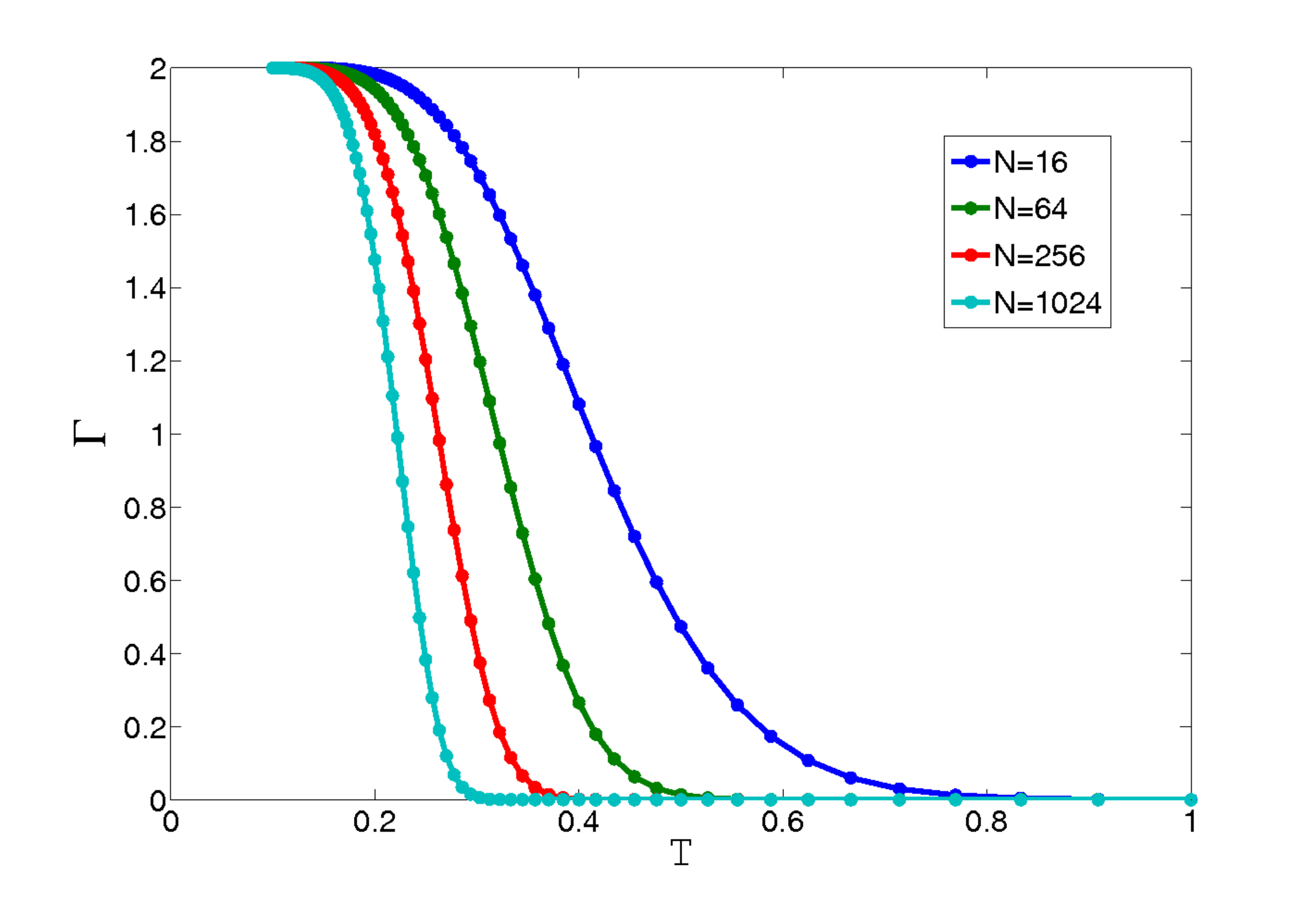}}
\caption{\label{fig2} A graph of $\Gamma$ against temperature $T=1/\beta$ for planar codes of various sizes $N$. Logarithms are taken base $2$. Topological order is found to be present for low temperature, but gets weaker for larger systems.}
\end{center}
\end{figure}

Let us characterize the size of a code by its total number of plaquettes, $N$. When calculating $\Gamma$ for systems of differently sized codes, the region $A$ should be chosen to grow with the system size. Let us choose to include half of all plaquettes within the region $A$, and hence $n_s = n_p = N/4$. The corresponding graphs of $\Gamma$ against inverse temperature $\beta$ are shown in Fig. \ref{fig2} for $J_S=J_P=1$. These graphs show remarkable agreement to those in previous studies for the topological entanglement entropy \cite{claudios,iblisdir}. In the thermodynamic limit and for any finite temperature it is easy to see that the anyonic occupation of $A$ will be completely random, giving $\Gamma = 0$. This reproduces the known result than the thermal state of the planar code is not topological ordered in this case \cite{claudios,iblisdir}.

\section{Possibilities for generalization}

The anyonic topological entropy has been defined here for use with Abelian anyon models, and claims about other cases are beyond the scope of this work. However we conclude with some discussion of how generalization to other cases may be possible. The definition of $\Gamma$ can be straightforwardly generalized to models for which point-like quasiparticles exist in a space of any dimension, rather than just two dimensions as above. The only requirements are than the region $A$ must be closed, the region $B$ must surround $A$, and the composite region $AB$ may not cover the entire model. The latter is required such that no global features (such as a conservation law) can be applied to deduce the net occupation of $A$ from $B$ even when error correction is not possible.

Notably, the parameter can be applied to cases for which the topological entanglement entropy cannot. The prime example is that of the planar code subject to periodic application of noisy stabilizer measurements, which gives a three dimensional (2 space and one time) configuration of defects \cite{dennis,raussendorf}. This model is the most realistic for quantum computation and has the highest known noise threshold, of around $1\%$ \cite{raussendorf}, and so the anyonic topological entropy may be useful for illuminating studies of this important model. 

Note that in other cases for spacial dimensions other than two, the phases detected by the parameter will not necessarily be topological in nature. This is because the above arguments concerning loop correlations may not apply. An example of the application of the parameter to a one-dimensional system is shown in the next section, where it detects the boundary of a ferromagnetic phase.

\section{Application to a one-dimensional model}

As an example of the calculation of $\Gamma$ for a one-dimensional model, we consider the 1D transverse field Ising model \cite{lieb}, which is equivalent to Kitaev's Majorana chain \cite{kitaev,laumann,loss}. The Ising model with no field can be considered to be a one-dimensional variant of the planar code, where the lattice used is a line rather than a square lattice as usual. With the field it is therefore a perturbed 1D planar code. The Hamiltonian of the model may be expressed,
\be
H_I = - w \sum_{j=1}^{N-1} \sigma^x_j \sigma^x_{j+1} + \frac{\mu}{2} \sum_{j=1}^{N} \sigma^z_j.
\ee
For $0 \leq \mu < 1/2$ it is known that the chain is in a ferromagnetic phase, which corresponds also to the ordered phase of the Majorana chain. For the case of $\mu=0$, the excitations of the Hamiltonian can be understood in terms of quasiparticles, which correspond to domain walls between regions of aligned spins. These can be used to calculate $\Gamma$. For $\mu>0$ these may not be the optimal choice of quasipartices, and so though $\Gamma >0$ is a witness of order, $\Gamma = 0$ is not evidence of its absence. This may therefore lead to us detecting the phase transition at a slightly lower value of $\mu$ than the true critical point.

\begin{figure}[t]
\begin{center}
{\includegraphics[width=8cm]{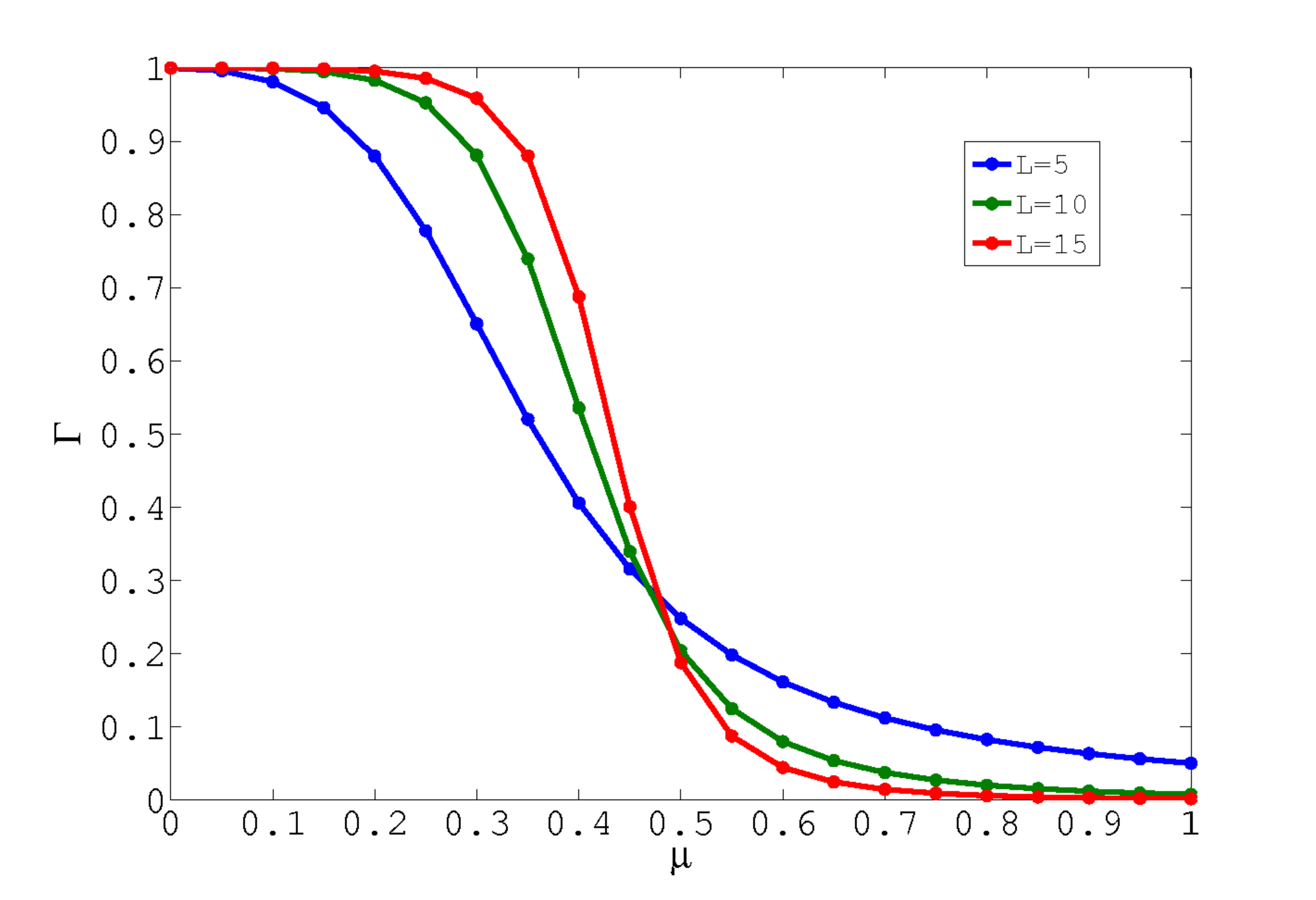}}
\caption{\label{fig3} A graph of $\Gamma$ against $\mu$ for $N=5$, $N=10$ and $N=15$. To calculate $\Gamma$ the chain of $N$ anyonic pseudo-spins was cut into five equal fifths. The first and fifth of these were used for region $C$, the second and fourth for $B$ and the third for $A$. $\Gamma$ here is calculated using logarithms of base 2.}
\end{center}
\end{figure}

To calculate $\Gamma$, we must first define a set of pseudo-spins whose states correspond to the occupations of the domain wall quasiparticles. $N-1$ such pseudo-spins are required, where the state of the $j$th pseudospin is defined as $\ket{0}$ if there is no domain wall between the $j$th and $j+1$th spins and $\ket{1}$ if there is. The states of these pseudo-spins contain all information about the relative orientations of the Ising spins, but not of the actual orientations themselves. To add this information, an $N$th pseudo-spin is introduced whose state is defined to be the same as that for the $N$th spin ($\ket{0}$ for spin up in the $x$-direction, $\ket{1}$ for spin down). The Hamiltonian $H_I$ can now be rewritten in the pseudo-spin space as follows,
\bq \nonumber
H_P = &-& w \sum_{j=1}^{N-1}  \sigma^z_j + \frac{\mu}{2} \left[\sigma^x_1 + \sigma^x_{N-1} \sigma^x_{N} + \sum_{j=1}^{N-2} \sigma^x_j \sigma^x_{j+1} \right].
\eq
With the Hamiltonian defined, the ground state may be calculated numerically for different values of $\mu$. The off diagonal terms may then be removed from the state, rendering it a classical distribution of domain walls, and the corresponding values of $\Gamma$ can be determined. The results of such a study are shown in Fig. \ref{fig3}. It is found that $\Gamma=\log2$ at $\mu=0$, showing that this state is ordered as expected, and $\Gamma = 0 $ for $\mu \gg 1/2$, showing that the order has been destroyed. Results for the system sizes considered intersect at around $\mu = 1/2$, suggesting that this is the transition point between the ordered and disordered phases, in agreement with known results. The intersection point appears to be just below the known critical value, which may be due to our use of $\mu = 0$ quasiparticles throughout.

It should be noted that a similar study could be performed for the toric code perturbed by magnetic fields. Though the phase transitions for this case have already been studied \cite{vidal}, the use of the anyonic topological would provide an independent verification of these results and may shed new light on the properties of the model. However, this would require a significant increase of the computation cost compared to the one-dimensional case, and so such a study is deferred to future work.

\emph{Acknowledgements:} The author would like to thank Daniel Loss, Alioscia Hamma, Diego Rainis and Abbas Al-Shimary for valuable discussions. This work was supported by the EPSRC and Swiss NSF.

\end{document}